\documentclass[prl,aps,preprint,showpacs,superscriptaddress]{revtex4}

\usepackage{graphicx}

\begin{document}

  \title{Novel Structural Motifs in Oxidized Graphene}
  
  \author{H. J. Xiang}
  %\thanks{Corresponding author. E-mail: hxiang@fudan.edu.cn}
  \affiliation{Key Laboratory for Computational Physical Sciences,
    Ministry of Education, P. R. China, and Department of Physics, Fudan
    University, Shanghai 200433, P. R. China}

  \author{Su-Huai Wei}
  \affiliation{National Renewable Energy Laboratory, Golden, Colorado 80401, USA}

  \author{X. G. Gong}
  \affiliation{Key Laboratory for Computational Physical Sciences,
    Ministry of Education, P. R. China, and Department of Physics,
    Fudan
    University, Shanghai 200433, P. R. China}

  \date{\today}

  \begin{abstract}
    The structural and electronic properties of oxidized graphene are
    investigated on the basis of the genetic algorithm and density
    functional theory calculations.
    We find two new low energy semiconducting phases of the fully oxidized
    graphene (C$_1$O).
    In one phase, there is parallel epoxy pair chains running along the zigzag
    direction. In contrast, the ground state phase with a slightly
    lower energy and a much larger
    band gap contains epoxy groups in three different ways: normal epoxy,
    unzipped epoxy, and epoxy pair. 
    %% Interestingly, the C$_1$O phase with the epoxy
    %% pair model has a lower conduction band minimum than the Dirac
    %% point of graphene. 
    For partially oxidized
    graphene, a phase separation between bare graphene and fully
    oxidized graphene is predicted.
    %% And the C$_1$O phase with the epoxy pair model can have
    %% a smooth low energy interface with bare graphene via a zigzag-like
    %% edge, but the resulting system is metallic with no local spin
    %% moment as a result of the vanishingly weak quantum confinement.
    \end{abstract}

  %61.48.Gh Structure of graphene
  %73.61.Wp Fullerenes and related materials
  %71.20.-b Electron density of states and band structure of crystalline
  %solids
  %73.22.-f Electronic structure of nanoscale materials and related systems

  \maketitle
Graphene, a single layer of honeycomb
carbon lattice, exhibits many exotic behaviors,
ranging from the anomalous
quantum Hall effect \cite{Novoselov2005,Novoselov2006,Zhang2005}  and
Klein paradox \cite{Katsnelson2006} to coherent transport
\cite{Miao2007}. 
Because of its exceptional electrical, mechanical, and thermal
properties, graphene  holds great promise for potential
applications in many technological fields such as nanoelectronics,
sensors, nanocomposites, batteries, supercapacitors and hydrogen
storage \cite{Geim2007}. 
Nevertheless, several of these applications are still not feasible
because the large-scale production of pure graphene sheets
remains challenging. 
Currently, graphene oxide (GO) is of
particular interest since chemical reduction
of GO has been demonstrated as a promising solution based
route for mass production of graphene
\cite{Eda2008,Gomez2007,Gijie2007,Tung2009,Li2008}. In addition, GO shows
promise for use in several technological applications such as polymer
composite \cite{Stankovich2006},
dielectric layers in nanoscale electronic devices, and the active region
of chemical sensors.

GO, which was first prepared by Brodie \cite{Brodie1859} in 1859, 
consists of oxidized graphene layers with
the hexagonal graphene topology \cite{Wilson2009}.
Although it is now widely accepted that GO bears hydroxyl and epoxide
functional groups on its basal plane \cite{Gao2009}, as confirmed by a recent
high-resolution solid-state $^{13}$C-NMR measurement  \cite{Cai2008}, the complete
structure of  GO has remained elusive because of the pseudo-random
chemical functionalization of each layer, as well as variations in
exact composition.

Several first-principles studies \cite{Kudin2008,Boukhvalov2008,Yan2009,Lahaye2009} have been
performed to investigate the structure and energetics of epoxide and hydroxyl
groups on single-layer graphene. In those calculations, 
some possible structures were manually selected and examined in order to obtain a good
structural model for GO. In this Letter, we adopt the genetic
algorithm (GA) in combination with density functional theory (DFT) to search the
global minimum structures of oxidized graphene with different
oxidation levels. As a first step, we consider only the arrangement of epoxide groups
on single-layer graphene.
We find two new graphene based low energy structures of C$_1$O. In both 
structures, there are epoxy pairs and some carbon-carbon $\sigma$
bonds are broken. However, the two structures of C$_1$O have dramatically
different electronic properties: one has a much larger band gap than
the other, and the low band gap structure has a conduction band
minimum (CBM) which is even
lower than the Dirac point of graphene. We also show that for
C$_x$O with low oxygen (high $x$) concentration, a phase separation 
between bare graphene and the low energy structures of C$_1$O will take
place in the ground state. In addition, the electronic structure of
C$_x$O will depend on which low energy structure is present in the system. 
Our results are directly relevant to the
oxidation of graphene in O$_2$ atmosphere \cite{Liu2008} or oxygen
plasma \cite{Kim2009}. They also have important
implications for understanding the structural and electronic
properties of GO made by mineral acid attack.

%GA details
We perform global structure searches based on 
the GA. The GA has been successfully applied to
find the ground state (GS) structure of clusters \cite{Deaven1995}, alloys
\cite{Liu2007}, and crystals \cite{Oganov2006}. In this work, we
consider all possible inequivalent (6 in total) graphene supercells up to 8 carbon
per cell. 
For each graphene supercell, we perform several GA
simulations to confirm the obtained GS structure. Here, the oxygen
atoms can only occupy the bridge sites of both sides of the graphene
plane. It can be easily seen that the number of
possible oxygen positions on both sides of the graphene is thrice the
number of carbon atoms. However, two oxygen atoms at nearest bridge
sites is very unstable due to the short O-O distances (about 1.2 \AA),
thus the lowest C/O ratio for a possible stable structure is 1. 
%% Therefore, the issue we are trying to address here is a fixed lattice
%% optimization problem. It is well known that the cluster expansion
%% approach \cite{Ferreira1989} established in the alloy theory was successfully used to
%% solve this kind of problems. 
%% As a matter of fact, the cluster expansion approach was adopted to
%% study the hydrogenation of graphene \cite{Xiang2009}.
%% However, the complexity (see discussion below) in the structure
%% of oxidized graphene makes the construction of the cluster expansion
%% Hamiltonian extremely difficult.
Different oxygen concentrations are simulated: C/O$=1$,
C/O$=2$, C/O$=3$ and C/O$=4$. 
For a given graphene supercell and C/O ratio, we first randomly
generate tens (e.g., 16) of structures of C$_x$O. Then we fully optimize the internal
coordinations and cell of structure. 
To generate new population, we perform the cut and splice crossover
operator \cite{Deaven1995,Oganov2006} on
parents chosen through the tournament selection. In an attempt to avoid stagnation and to
maintain population diversity, a mutation operation in which
some of the oxygen atoms can hop to other empty bridge sides is introduced.
In this study, we use DFT to calculate the energies and relax the
structures. Our first principles DFT calculations are performed on the
basis of the projector augmented wave method \cite{PAW} encoded in
the Vienna ab initio simulation package \cite{VASP} using the local density
approximation (LDA) and the plane-wave cutoff energy of 500 eV.

\begin{figure}
    \includegraphics[width=7.5cm]{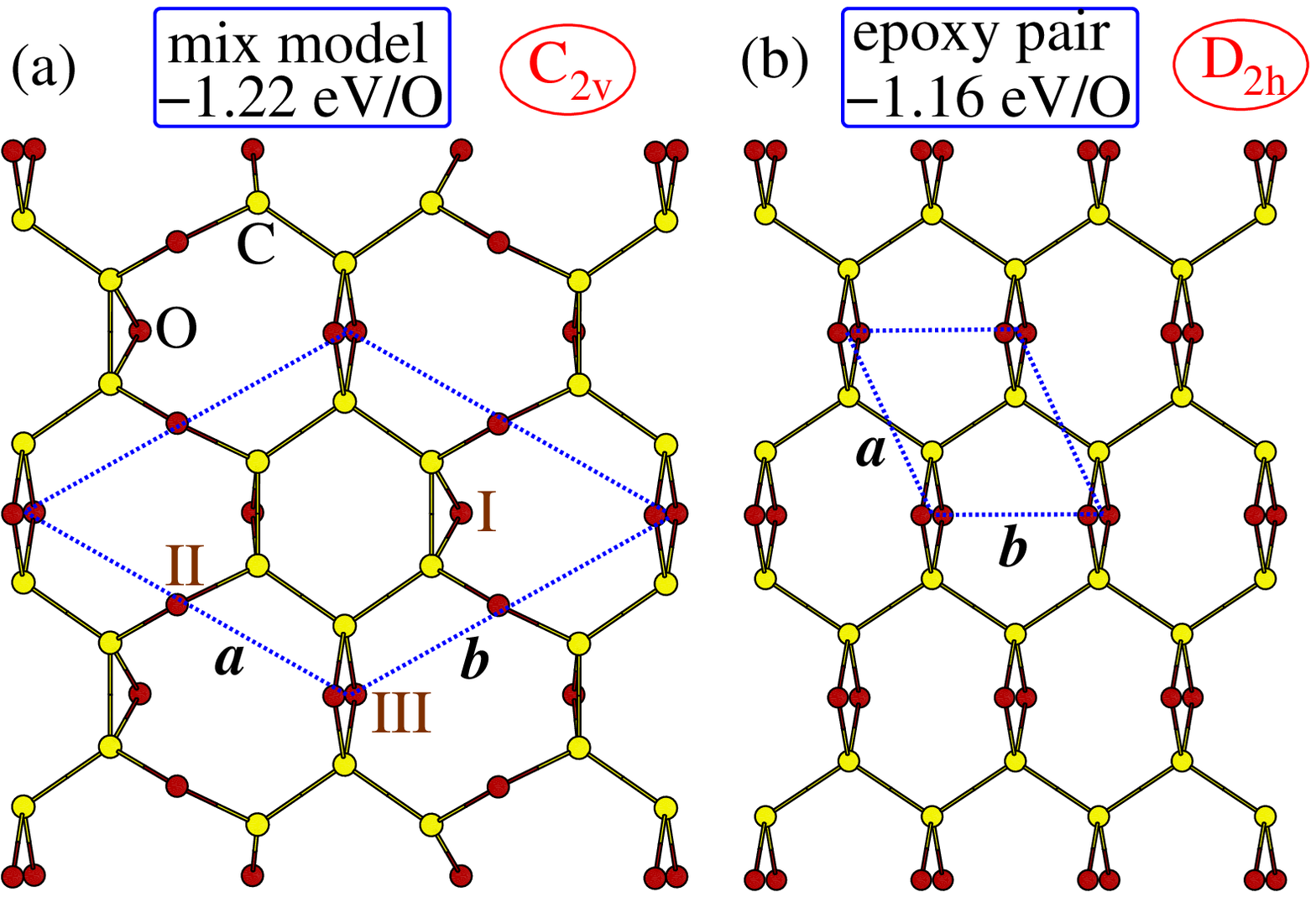}
    \caption{ (Color online) (a) The structure of the mix model
      ($C_{2v}$ symmetry) which
      is predicted to be the lowest energy C$_1$O phase.
      (b) The low energy C$_1$O phase with the epoxy pair model
      ($D_{2h}$ symmetry), which has
      a higher energy by only 0.06 eV/O than the mix model.
      ``I'', ``II'', and ``III'' in (a) indicate normal epoxy,
      unzipped epoxy, and epoxy pair, respectively. 
      The unit cells are enclosed by dashed lines.
      The numbers give the formation energy of the oxidized graphene
      structures. }
    \label{fig1}
  \end{figure}

%CO,C2O,C3O,C4O and phase separation
To characterize the stability of oxidized graphene structures, we define the
formation energy or oxygen absorption energy as:
\begin{equation}
  E_f = E({\mathrm C}_x{\mathrm O}) - x \mu_{\mathrm C} - \mu_{\mathrm
    O}\, 
\end{equation}
where $\mu_{\mathrm C}$ is set to be the energy of graphene per carbon
atom, and
$\mu_{\mathrm O} = 1/2 E({\mathrm O}_2)$ [$E({\mathrm O}_2$) is the
energy of an isolated triplet oxygen molecule]. It should be noted that
the relative stability between various structures of oxidized graphene
does not depend on the particular choice of the oxygen chemical potential ($\mu_{\mathrm
  O}$). 
Our GA simulations reveal two low energy structures of C$_1$O, as 
shown in Fig.~\ref{fig1}. In both structures, each of the carbon atoms
bonds with two neighbor carbon atoms and two O atoms, thus there is no
$sp^2$ carbon.  
The structure shown in Fig.~\ref{fig1}(b) is rather simple: 
There are epoxy pair chains along the zigzag direction which are parallel to each other. 
We will refer to this model as the epoxy pair model ($D_{2h}$ symmetry) hereafter.
The formation of an isolated epoxy pair was suggested in a theoretical
study on the graphene oxidative process \cite{Li2009}.
It was also shown that 
an isolated epoxy pair is less stable than a carbonyl pair
\cite{Li2009}.
However, in the fully oxidized graphene case, we find that the epoxy
pair chain is more stable than the carbonyl pair chain by about 0.84 eV per
pair.
The lowest energy structure ($C_{2v}$ symmetry) [Fig.~\ref{fig1}(a)] of C$_1$O has a lower
energy than the epoxy pair model by only 60 meV/O. The small energy
difference suggests that both phases might coexist at finite temperature.
In the lowest energy structure, 
there are isolated six-membered carbon rings, which are connected to
the neighboring six-membered carbon rings through two epoxy pairs and
four unzipped epoxy groups. For each six-membered carbon ring, there are
two normal epoxy groups.     
Therefore, we term this complex structure as the mix model.
One  can obtain the epoxy pair model from the mix model by moving the
unzipped epoxy groups on top of the normal epoxy groups.

\begin{figure}
    \includegraphics[width=9.5cm]{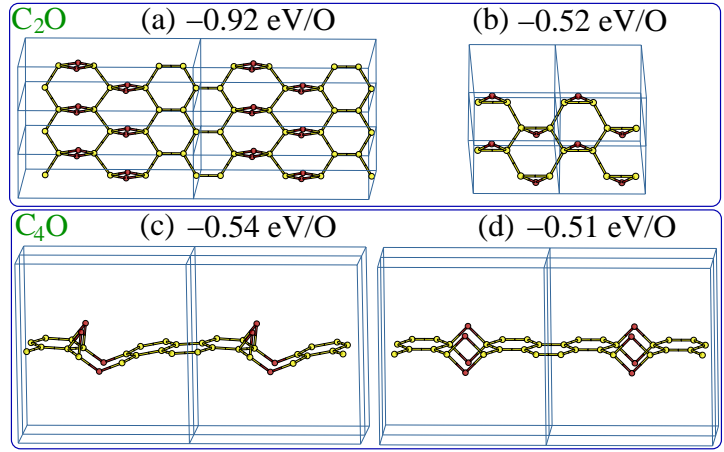}
    \caption{ (Color online)
      (a) shows the newly predicted lowest energy structure of the C$_2$O
      phase among all configurations with no more than eight carbon
      atoms per cell.
      (b) is the model of the C$_2$O phase predicted by 
      Yan {\it et al.} \cite{Yan2009}. 
      (c) shows the lowest energy structure of the C$_4$O
      phase among all configurations with no more than eight carbon
      atoms per cell. 
      (d) is a low energy structure of the C$_4$O
      phase  which has a higher energy by only 0.03 eV/O than the
      lowest energy structure. }
    \label{fig2}
  \end{figure}

The low energy structures of oxidized graphene with C/O$=2$ and
C/O$=4$ are displayed in Fig.~\ref{fig2}.
Among all C$_2$O structures with no more than eight C atoms per cell, 
the lowest energy structure has two neighboring epoxy pair chains and
half of the carbon atoms remain $sp^2$ hybridized.
For comparison, we also show the C$_2$O structure predicted by Yan
{\it et al.} \cite{Yan2009} in which no carbon-carbon bond is broken and all carbon
atoms are $sp^3$ hybridized. Test calculations show that our
identified new structure is more stable by almost 0.40 eV/O than Yan's structure with
normal epoxy groups. For C$_4$O, we find that oxygen atoms tend to
form isolated chains. In the lowest energy structure, the graphene is
unzipped by an oxygen chain and the normal epoxy groups are connected with
the unzipped epoxy groups. In another metastable structure, there is
an isolated epoxy pair chain. The isolated epoxy pair chain structure
is less stable than the lowest energy structure by only 30 meV/O.
The isolated epoxy pair chain structure is found to be the 
lowest energy structure of C$_3$O with no more than six carbon atoms.

The calculated formation energies of
the above discussed lowest energy structures are about $-1.2$ eV/O,
$-0.9$ eV/O, and $-0.5$ eV/O for C$_1$O, C$_2$O, and C$_4$O,
respectively. We can clearly see that C$_1$O has the lowest formation
energy and the formation energy increases with
the decrease of the oxygen concentration. 
We note that the above results are based on the GA simulations with finite
supercell size. We now discuss the possible structure of oxidized graphene with an
infinite large graphene supercell. 
The two low energy structures of C$_1$O
should remain the same because of the perfect ordering and low
formation energy.  For oxidized graphene with less oxygen atoms than
carbon atoms, the phase separation between the C$_1$O phase and bare
graphene will occur. The phase separation is caused by the tendency to
minimize the number of broken carbon-carbon $\pi$-$\pi$ bonds, as in
the case of partially hydrogenated graphene \cite{Xiang2009}. The
tendency toward phase separation is manifested in the lowest energy
structure of C$_2$O [Fig.~\ref{fig2}(a)] and is confirmed in the case
of C$_4$O: If we double the 
cell of Fig.~\ref{fig2}(d) along the direction perpendicular to the
isolated epoxy pair chain and move the two epoxy pair chains close to
each other, the resulting C$_4$O structure will have a formation
energy of $-0.87$ eV/O. In contrast,  increasing the cell of
Fig.~\ref{fig2}(c) does not lead to a significant decrease of the
formation energy.

\begin{figure}
    \includegraphics[width=7.5cm]{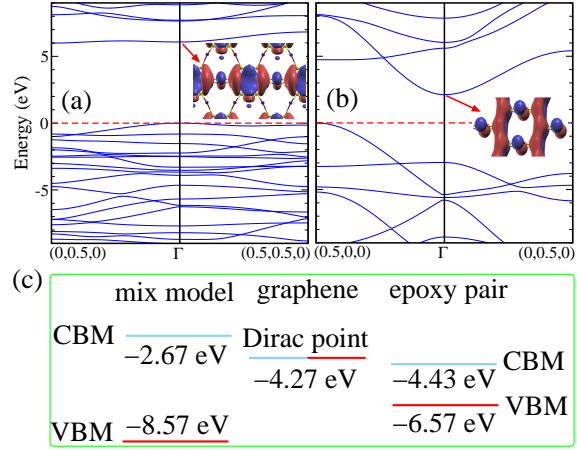}
    \caption{ (Color online)
      (a) and (b) show the band structures from the hybrid HSE06
      calculations of the C$_1$O phases with the
      mix model and epoxy pair model, respectively.  
      In the insets, we show the wavefunction plots of the LUMO states
      at $\Gamma$. The k-points are given in terms of the reciprocal lattice.
      (c) Schematic illustration of the band alignment between the two C$_1$O phases and
      graphene from the hybrid HSE06 calculations. }
    \label{fig3}
  \end{figure}

%band structure and band offset of C$_1$O
Our calculations indicate that 
the two low energy phases of C$_1$O will probably 
exist in any oxidized graphene.
Therefore, the electronic properties of the C$_1$O phases is of paramount
importance.  Because the LDA is well known to underestimate the band
gaps of semiconductors, we calculate the band structure of the C$_1$O phases
by employing the screened
Heyd-Scuseria-Ernzerhof 06 (HSE06) hybrid functional
\cite{Heyd2003},  which was shown to give a good band gap
for many semiconductors including graphene based systems \cite{Hod2008}. 
Our calculations show that the C$_1$O phase with the mix model is a
semiconductor with a large band gap of 5.90 eV
[Fig.~\ref{fig3}(a)]. The valence band maximum (VBM) 
locates at $\Gamma$, but the CBM is almost
non-dispersive, which is consistent with the fact that 
the CBM state at $\Gamma$ is an antibonding C-O orbital mostly
localized in the six-membered carbon rings [see the inset of Fig.~\ref{fig3}(a)].
Fig.~\ref{fig3}(b) shows that the C$_1$O phase with the epoxy
pair model is also a semiconductor but with a smaller indirect band
gap of 2.14 eV. The VBM and CBM locate at (0.5, 0, 0) and $\Gamma$,
respectively. An important difference between the two phases is that
the C$_1$O phase with the epoxy pair model has a much smaller electron and
hole effective mass. This is because the band overlap is stronger
in the epoxy pair model with a higher symmetry, as can be seen from
the CBM wavefunction displayed in the inset of Fig.~\ref{fig3}(b).  
The larger band gap and lower energy in the mix model is 
a consequence of its lower symmetry ($C_{2v}$):
Level repulsions between the occupied and unoccupied
bands which are symmetrically forbidden in the epoxy pair model ($D_{2h}$) become
possible in the mix model.

We also calculate the absolute values of the VBM and CBM levels using
the HSE06 hybrid functional, which was found
\cite{Alkauskas2008,Xiang2009B} to give improved results compared to
the LDA functional. The vacuum level is
defined as the average electrostatic potential in the vacuum
region where it approaches a constant \cite{Xiang2009B}. 
For comparison, we also calculate the work function of graphene.
Our results are shown in Fig.~\ref{fig3}(c). As expected, the VBM
of both C$_1$O phases are below the Fermi level of
graphene due to the oxygen $2p$ orbitals with low energies. And the CBM level
of the C$_1$O phase with the mix model is 
higher than the Fermi level of graphene by 1.6 eV.
Interestingly and surprisingly, the CBM level of the C$_1$O phase with
the epoxy pair model is lower than the Dirac point of graphene. This
is because the low oxygen 2s and 2p orbitals and the formation of one
dimensional zizag chain by carbon 2p$_z$ orbitals dramatically lowers
the eigenvalue of the CBM state [see the 
  inset of Fig.~\ref{fig3}(b)]. 
It is noted that LDA gives qualitatively similar band alignment.
This will have interesting consequences as will be discussed below.

\begin{figure}
    \includegraphics[width=9.5cm]{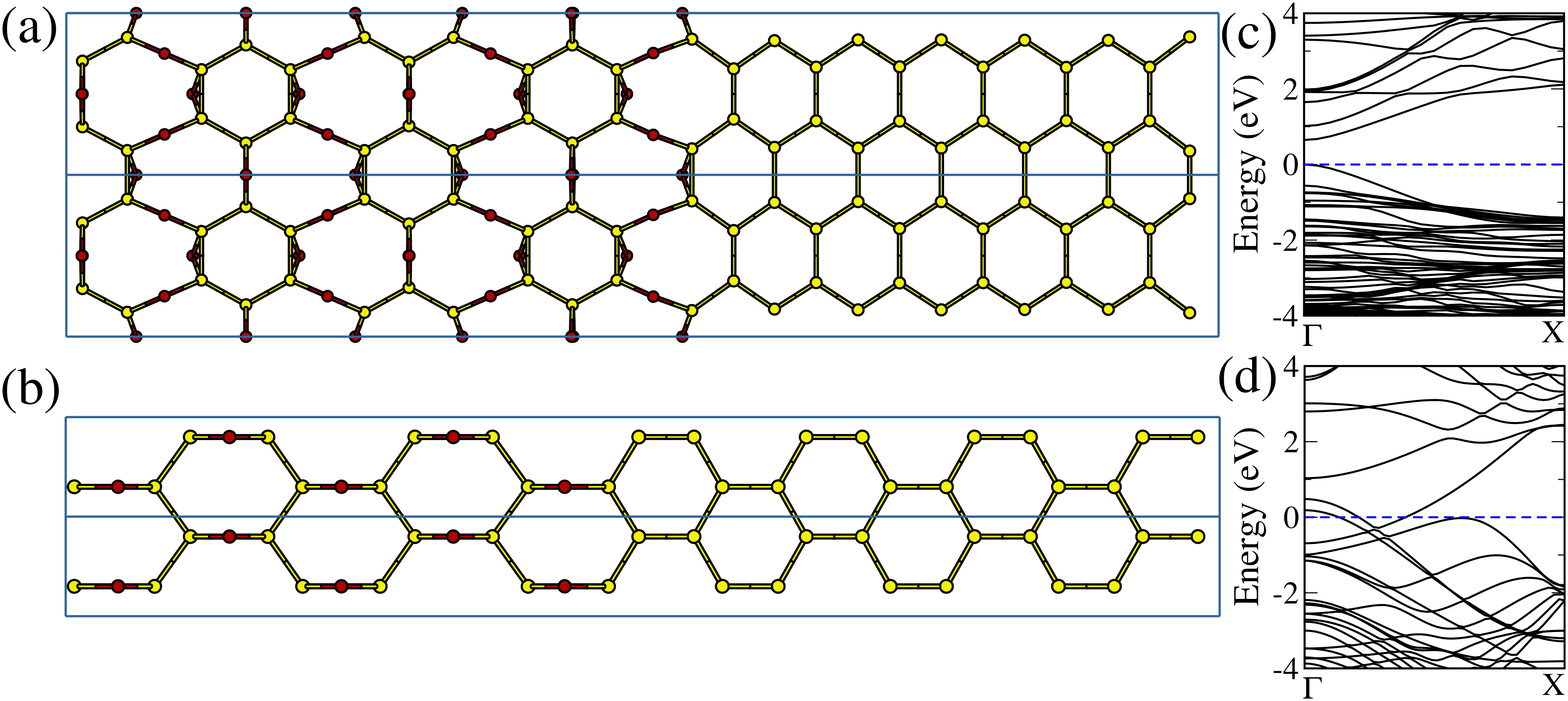}
    \caption{ (Color online)
      (a) shows a partially oxidized graphene with an armchair edge 
      between bare graphene and the C$_1$O phase with the mix model.
      (b) displays a partially oxidized graphene with a zigzag-like edge
      between bare graphene and the C$_1$O phase with the epoxy pair
      model.  
      (c) and (d) show the corresponding LDA band structures.
    }
    \label{fig4}
  \end{figure}

%GNR created by C$_1$O pattern
The experimentally synthesized graphene oxide usually has a much lower
oxygen concentration than the C$_1$O phases. Here, we will discuss the
electronic structures of partially oxidized graphene. 
As shown above, there is a phase separation between the graphene part
and the C$_1$O phase in the partially oxidized graphene under
thermodynamic equilibrium. 
The C$_1$O phase with the epoxy pair model could match graphene well via a
zigzag-like edge with a lattice mismatch of 6.0 \%
[Fig.~\ref{fig4}(b)].  
However, the lattice mismatch between the C$_1$O phase with the mix model
and graphene is much larger: The smallest mismatch (21.9\%) occurs  when the C$_1$O
phase connects with  graphene via an armchair edge [Fig.~\ref{fig4}(a)]. 
As expected, there is a band gap openning [See Fig.~\ref{fig4}(c)] in
the partially oxidized 
graphene with a phase separation between graphene and the C$_1$O phase
with the mix model. This is due to quantum confinement effect, similar
to the armchair graphene nanoribbon case \cite{Yang2007}.
In contrast, the partially oxidized graphene with a zigzag-like edge
between graphene and the C$_1$O phase with the
epoxy pair model is metallic [See Fig.~\ref{fig4}(d)]. 
And there is no
flat band and inclusion of spin degree of freedom does not open a gap, different
from the zigzag graphene nanoribbon case \cite{Yang2007}.
The metallicity of the system is not due to the peculiar zigzag-like
interface between graphene and the C$_1$O phase because a test
calculation on a system similar to that shown in Fig.~\ref{fig4}(a)
but with an armchair interface between graphene and the C$_1$O phase
with the epoxy pair model is also metallic.
This is due to the fact that the CBM of the C$_1$O
phase with the epoxy pair model is lower than graphene, thus there is
almost no quantum confinement effect and no flat edge states.  

%summary

  In summary, we have performed global search of the lowest energy
  structures of oxidized graphene using the genetic algorithm approach
  combined with density functional theory.
  Our calculations unravel two novel low energy semiconducting phases of fully oxidized
  graphene C$_1$O. 
  The C$_1$O phase with the epoxy pair model has an indirect band gap
  of about 2.14 eV, and an extremely low CBM that is below the Fermi level
  of graphene. The ground state of the C$_1$O phase with three
  mixed epoxy groups has a lower energy
  by only 60 meV/O than the C$_1$O phase within the epoxy pair model
  and a much larger band gap. 
  Our calculations predict that the phase separation between bare
  graphene and fully oxidized graphene is thermodynamically favorable
  in partially oxidized graphene. The C$_1$O phase with the epoxy pair
  model has a much smaller lattice mismatch with graphene than the
  case of the mix model.
  In the partially oxidized graphene with a zigzag-like edge between
  graphene and the C$_1$O phase with the epoxy pair model, there is no
  band gap openning and no local spin
  moment as a result of the unusually low CBM of the C$_1$O phase.
  
  %%Acknowledgements
  Work at Fudan was supported by the National Science Foundation
  of China.
  Work at NREL was supported by the U.S. Department of Energy, under
  Contract No. DE-AC36-08GO28308.
  We thank Dr. Gus Hart for useful discussion at the early stage on this
  project.

  \clearpage

\end{document}